\documentclass[11pt,twoside]{article}
\usepackage{asp2010}

\begin{document}

\title{Environmental effects in the interaction and merging of galaxies \\in zCOSMOS}
\author{P. Kampczyk, S. J. Lilly and the zCOSMOS Collaboration
\affil{Institute for Astronomy, ETH Zurich, \\Wolfgang-Pauli-strasse 27, 8093 Zurich, Switzerland}
}

\begin{abstract}
The zCOSMOS-bright 10k spectroscopic sample reveals a strong environmental dependence of close kinematic galaxy pair fractions in the studied redshift range $0.2<z<1$. The fraction of close pairs is three times higher in the top density quartile than in the lowest one. This environmental variation in pair fractions will translate into merger fractions since merger timescales are shown, based on Millennium simulation catalogs, to be largely independent of environment, once accounted for projection effects. While galactic properties of close kinematic pairs (morphologies and star formation rates) may seem to be non-representative of an underlying, global galaxy population, they can be explained by taking into account well-known effects of environment, and changes caused by galaxy interactions. The latter is responsible for an increase of irregular galaxies in pairs by a factor of 50-75\%, with a disproportionate increase in the number of irregular-irregular pairs (4-8 times), due to disturbance of about 15\% of the disk galaxies in pairs. Another sign of interaction is an observed boost in specific star formation rate (factor 2-4) for the closest pairs. While significant for paired galaxies, this triggered star-formation due to interactions represents only about 5\% of the integrated star-formation activity in our volume-limited sample. Although majority of close kinematic pairs are in dense environments, the effects of interactions appear to be strongest in the lower density environments. This may introduce strong biases into observational studies of mergers, especially those based on morphological criteria. Relative excess of post-starburst galaxies observed in paired galaxies (factor $\sim$2) as well as excess of AGNs (factor of over 2), linked with environmental dependence of the pair fractions could indicate that early phases of interactions and merging are plausible candidates for environmental quenching, observed in the global galaxy populations.
\end{abstract}

\section{Introduction}
Recent local and intermediate redshift surveys have provided observational evidence of enhanced fractions of close kinematic pairs of galaxies in intermediate to high-density regions \citep{McIntosh2008,Darg2010,Perez2009,Ellison2010,Lin2010,deRavel2010,Kampczyk2011}. We investigate, whether those observed fractions will translate into a similar environment dependent merger rates, indicating a faster build-up of mass in denser environments via merging. We also study properties of galaxies in pairs, which can be explained as a combination of environmental effects, pre-shaping future pair populations, and changes in galaxy properties due to the interactions in close pairs. The latter is discussed also as a function of environment, indicating observational biases for morphology-based merger studies.

\section{Data}

We use the redshift information recently derived in the zCOSMOS-bright project \citep{Lilly2007,Lilly2010} Ð a major spectroscopic redshift survey of the galaxies in the COSMOS field \citep{Scoville2007}. The "10k sample", we operate on, is based on the data obtained in the first two observing seasons. It yields spectra of over 10,000 galaxies with $I_{AB}<22.5$ across $1.7$ deg$^{2}$ of the COSMOS field. With high success rate in measuring redshifts (close to 100\% at $0.5<z<0.8$), high angular completeness up to small scales, and velocity accuracy of $110$ kms$^{-1}$ zCOSMOS is well suited for study of close kinematic pairs of galaxies and their environments.

	Using secure redshifts and absolute magnitude cut $M_{B,AB} < -19.64 -1.36z$ yields a volume-limited  sample of 3667 galaxies up to $z = 1$ ("global sample"). From these, the close kinematic pair galaxies are being selected with the following criteria: velocity difference $dv < 500$ kms$^{-1}$ and a projected proper (physical) separation $dr <100$ h$^{-1}$kpc i.e.~$140$ kpc for our chosen $h = 0.7$, or sub-samples selected to have smaller dr. Additionally we require that the galaxies in pairs are major merger progenitors Ð i.e.~ the difference of their $M_B$ absolute magnitudes should not be larger than $1.5$ mag. These selection criteria yield 153 close kinematic pairs up to redshift $z = 1$.
Detailed pair statistics as a function of redshift, as well as details of completeness corrections are described in our other work - \citet{deRavel2010}.

For the purpose of environmental studies, the three-dimensional density field has been reconstructed utilizing both spectroscopic (10k) and photometric redshifts (30k) for the full flux-limited sample in \citet{Kovac2010}. We use those robust over-density estimates to study environmental dependencies of galaxy pair fractions and properties.

	\citet{Knobel2009} have generated from the 10k sample a superior hybrid group catalog, which consists of 800 identified groups of richness 2 or more. Except for very low redshift, all pairs will satisfy selection criteria of the group catalog.

In this analysis we use the structural parameters and automated morphological classification based on the classification scheme derived by \citet{Scarlata2006}.
Spectra derived in zCOSMOS 10k sample in the redshift range $0.5<z<0.9$ give measures of emission line fluxes for OII. These were used to obtain star-formations rates SFR and specific star formation rates sSFR - see \citet{Maier2009}.

\section{Environmental dependencies}

In the analysis of environments of galaxy pairs we are using a stellar mass-weighted density measure that is computed on a fixed aperture of $3$ Mpc (comoving) radius that is much larger than the separations of the pairs, and velocity offset of 1000 kms$^{-1}$. This estimator has the advantage of being, in principle, preserved throughout the process of merging.

	We correct for chance alignments that enable physically unrelated galaxies to fulfill our pair selection criteria. The number of such "interlopers" clearly increases as the local projected density of galaxies increases. For the highest over-density quartile and the largest projected separation $dr < 100$ h$^{-1}$kpc this correction decreases number of observed pairs by 23\%. It is more modest for smaller projected separations and essentially negligible for the range below the median of the density field.

	Calculated pair-fractions, corrected for projection effects, rise towards higher over-densities. The lowest quartile (defined by the global galaxy sample) of the over-densities contains only 10\% of the pair galaxies, and the fraction of pair galaxies in the highest over-density quartile is typically 2-3 times higher than in the lowest quartile for all of the $dr$ bins considered: $30, 50$ and $100$ h$^{-1}$kpc.

	This strong environmental dependence of pair fractions will translate into a higher merger rates in denser environments, under assumption that the fraction of close pairs that will merge in a given timescale, will be independent of environment. As we showed in \citet{Kampczyk2011}, based on our analysis of the Millennium simulation mocks \citep{Kitzbichler2008}, merger timescales are indeed largely independent of environment, once accounted for projection effects.

	The properties of galaxy populations vary with environment. Since the paired galaxies are preferentially drawn from more over-dense regions, we must carefully take this into account, while creating a proper ÒparentÓ comparison samples.

\section{Impact of environmental dependencies and interactions}
The morphologies of galaxies in the close kinematic pairs are not representative of the global galaxy sample, but rather reflect the morphological mix of the richer environments that they reside in. The fraction of spheroidal types in pairs is higher than that for field galaxies, but is exactly the same as for the population of group galaxies.

	Monte Carlo simulations used to create comparison samples of the same redshift distributions and environments (i.e.~group galaxies) as paired galaxies enabled us to match well the abundances of spheroidal galaxies in pairs. However, the fraction of irregulars in pairs is significantly elevated (50-75\%), at the expense of disk galaxies ($\sim$15\%). This is presumably due to interactions in pairs that perturb morphologies of disk galaxies involved. The combinations of morphologies in individual close kinematic pairs also show an increase in the relative number of irregular - irregular pairs (4-8 times) indicating that morphological disturbance in one galaxy is usually accompanied by a disturbance of the other. In fact, 60\% of the irregular galaxies in close kinematic pair sample with $dr < 50$ h$^{-1}$kpc have an irregular companion, even though irregulars comprise only 15\% of that sample. The disk - disk pairs are the dominant type in our sample, and spheroid - spheroid pairs, which would be expected to be future "dry" merger systems, are the least abundant morphological combination, contributing on average only $\sim$5\% of the close kinematic pair systems in the redshift range $0.2<z<1$. This fraction probably increases towards low redshift due to the increase of the early-type population.
	
	Galaxies in close kinematic pairs show an enhancement of SFR and sSFR that increases with decreasing projected separation, indicative of induced star formation. The excess in the sSFR at $dr < 30$ h$^{-1}$kpc, is about a factor of two. Coupled with the overall pair fraction (corrected for spatial sampling), this implies that close kinematic pairs are contributing about 10\% of the total star-formation seen in our volume limited galaxy sample, about a half of which can be attributed to "excess" star - formation associated with the interaction.

\section{Post-starburst galaxies and AGNs in close kinematic pairs}
Post-starburst galaxies identified by \citet{Vergani2010} are two times more abundant among close kinematic pair galaxies than for the global sample. This, together with an evidence for triggered star formation, indicates that, at least in some cases, galaxy interactions in pairs can also be responsible for closing down star-formation.

	In our analysis of AGN fractions of paired galaxies \citep{Silverman2011} we find an enhancement of a factor greater than two, while compared to non-paired galaxies, suggesting that  interactions and early phases of merging can trigger AGN activity.

 	Interactions in pairs seem to have an influence on the properties of global galaxy populations and this effect does depend on environment. This suggests that interactions in pairs and subsequent merging might be a viable candidate for Òenvironmental quenchingÓ as defined in \citet{Peng2010}.

\section{Environmental biases}

Interestingly, the interaction-induced activity is preferentially seen in pair galaxies at close separations in \emph{low-density} environments, where the boost in sSFR can reach 2-4 times and fraction of galaxies with disturbed morphologies is highest. In above-average over-dense environments, which comprise the majority of the close kinematic pair systems, galaxies do not show strong signatures in their distributions of asymmetries, sSFR, or irregular fractions when compared with suitable parent galaxy samples. 

	Studies based on identifying galaxies or pairs with clear signs of interactions may therefore be biased to lower density environments, sampling only a minority of kinematic pairs. The severity of this bias may be redshift dependent. We might expect these effects to produce steeper redshift evolution of disturbed object fractions than in studies based on close kinematic pair systems. This may be the cause of the wide discrepancies in the literature about the redshift evolution of the merger rate.

	This adds to the list of possible observational biases, which we indicated in our morphology - based studies of COSMOS mergers in \citet{Kampczyk2007}.

\bibliographystyle{asp2010}

\bibliography{kampczyk}

\end{document}